\documentclass[twocolumn]{aastex62}
\usepackage{tablefootnote}
\usepackage{threeparttable}
\usepackage{tabularx} 
\usepackage{txfonts}
\usepackage{color}

\shorttitle{On the origin of BH spin in HMXBs}
\shortauthors{Qin et al.}

\begin{document}

\title{On the Origin of Black Hole Spin in High-mass X-Ray Binaries}

\author{Ying Qin}
\affiliation{Geneva Observatory, University of Geneva, CH-1290 Sauverny, Switzerland}
\affiliation{Center for Interdisciplinary Exploration and Research in Astrophysics (CIERA) and Department of Physics and Astrophysics,
Northwestern University, Evanston, IL 60208, USA}

\author{Pablo Marchant}
\affiliation{Center for Interdisciplinary Exploration and Research in Astrophysics (CIERA) and Department of Physics and Astrophysics,
Northwestern University, Evanston, IL 60208, USA}

\author{Tassos Fragos}
\affiliation{Geneva Observatory, University of Geneva, CH-1290 Sauverny, Switzerland}

\author{Georges Meynet}
\affiliation{Geneva Observatory, University of Geneva, CH-1290 Sauverny, Switzerland}

\author{Vicky Kalogera}
\affiliation{Center for Interdisciplinary Exploration and Research in Astrophysics (CIERA) and Department of Physics and Astrophysics,
Northwestern University, Evanston, IL 60208, USA}

\begin{abstract}
Black hole (BH) spins in low-mass X-ray binaries (LMXBs) cover a range of values that can be explained by accretion after BH birth. In contrast, the three BH spin measurements in high-mass X-ray binaries (HMXBs) show only values near the maximum and likely have a different origin connected to the BH stellar progenitor. We explore here two possible scenarios to explain the high spins of BHs in HMXBs: formation in binaries that undergo mass transfer (MT) during the main sequence (MS; Case-A MT), and very close binaries undergoing chemically homogeneous evolution (CHE). We find that both scenarios are able to produce high-spin BHs if internal angular momentum (AM) transport in the progenitor star after its MS evolution is not too strong (i.e., weak coupling between the stellar core and its envelope). If instead efficient AM transport is assumed, we find that the resulting BH spins are always too low with respect to observations. The Case-A MT model provides a good fit for the BH spins, the masses of the two components, and the final orbital periods for two of the three BHs in HMXBs with measured spins. For one of them, the mass predicted for the BH companion is significantly lower than observed, but this depends strongly on the assumed efficiency of mass transfer. The CHE models predict orbital periods that are too large for all three cases considered here. We expect the Case-A MT to be much more frequent at the metallicities involved, so we conclude that the Case-A MT scenario is preferred. Finally, we predict that the stellar companions of HMXBs formed through the Case-A MT have enhanced nitrogen surface abundances, which can be tested by future observations.
\end{abstract}


\section{Introduction} 
X-ray binaries are a class of binary stellar systems containing a compact stellar remnant, either a neutron star or a black hole (BH), accreting from a non-compact companion (donor) star. X-ray binaries are often divided into high-mass X-ray binaries (HMXBs) or low-mass X-ray binaries (LMXBs) according to the mass of the donor star. While in LMXBs the donor star overfills its Roche lobe, transferring mass to the compact object through the first Lagrangian point, HMXBs are most often wind-fed systems, where the compact object is capturing and accreting part of the strong stellar wind of its massive donor star companion. Interestingly, all three dynamically confirmed BH HMXBs have massive main-sequence (MS) companion stars (see Table 1), in a few-day orbits, where the companion is close to filling its Roche lobe \citep [Roche-lobe filling factors $>80\%$, see] []{2011ApJ...742...84O,2014MNRAS.440L..61Z}.

We should note that some types of BH X-ray binaries, like the BH HMXB candidates IC10 X-1 and NGC300 X-1, are potential progenitors of double BHs \citep{2011ApJ...730..140B}. However, these two systems have Wolf-Rayet companion stars, and the measured velocities are most likely due to the stellar winds of the BH companion instead of its orbital motion \citep{2015MNRAS.452L..31L}, which makes the dynamical measurement of the BH mass unreliable.

Over the last decade, the BH spins of 20 X-ray binaries \citep[][and references therein]{2015PhR...548....1M} have been measured using two main methods: the continuum fitting method \citep[][and references therein]{2014SSRv..183..295M} and the iron (Fe) K$\alpha$ line fitting method \citep[][and references therein]{2014SSRv..183..277R}. For LMXBs, the measured spins (namely, $a_* \equiv cJ/GM^2$, where J and M are the AM and mass of the star, $c$ is the speed of light, and $G$ is the gravitational constant) of BHs span the entire range from zero to maximally spinning. Based on the standard isolated binary formation channel, the origin of the BH spin in these binaries can be explained through accretion onto the BH after its birth \citep{2003MNRAS.341..385P,2015ApJ...800...17F,2017A&A...597A..12S}.

In contrast, all three of the BH spins measured in HMXBs have been found to be near maximal (see Table 1). Accretion after BH formation was also proposed to explain such a high spin \citep{1994ApJ...436..843B,2011MNRAS.413..183M}, but the lifetime of the massive companion star was too short \citep{2010Natur.468...77V,2012ApJ...747..111W} to significantly spin up the BH assuming Eddington limited accretion. Hence, it would require significantly super-Eddington (mass transfer) MT rates for the BH to accrete any appreciable amount of material. Furthermore, it is unclear how a wind-fed system with a MS accretor can reach such high MT rates, and there is no observational evidence that either of the three observed BH HMXBs are currently undergoing super-Eddington MT.
Most recently, it was suggested that slow ejecta from a failed supernova that formed the BH can interact with the companion and be torqued, increasing their specific angular momentum (AM) before falling back onto the newly formed BH \citep{2017ApJ...846L..15B}. However, follow-up simulations showed that realistic velocity profiles of the supernova ejecta can only lead to mild spin-up of the BH \citep{2018ApJ...862L...3S}. Alternatively, it has been suggested that gravity waves during the very last phases of the evolution of massive stars \citep{2015ApJ...810..101F} or instabilities during the core collapse phase \citep{2016NewA...44...58M} can add AM in the collapsing core in a stochastic way. But in both cases the amount of AM that can be transferred cannot lead to a significant BH spin.

\begin {table*}[t]
\caption {Main Properties of High-mass X-Ray Binaries with Measured Spins.}
\centering
\setlength{\tabcolsep}{8.0mm}{
\begin{tabular}{ l l  l  l  l  l l}
\hline \hline
Sources &$M_1/M_\odot$ &$M_2/M_\odot$ &$a_*$ & $P$/days &References \\
M33 X-7 &$15.65\pm{1.45}$ &$70.0\pm{6.9}$ &$0.84\pm{0.05}$ &3.45  &(1),(2),(5) \\
Cygnus X-1 &$14.8\pm{1.0}$ &$19.16\pm{1.90}$ &$>$0.983 &5.60 &(6),(7) \\          
LMC X-1 &$10.9\pm{1.4}$ &$31.79\pm{3.67}$ &$0.92_{-0.07}^{+0.05}$ &3.91 &(3),(4) \\
\hline 
\end{tabular}}
  \begin{tablenotes}{ 
  \item
  \textbf{References}: (1) \citealt{2007ATel..977....1O}, (2) \citealt{2008ApJ...679L..37L}, (3) \citealt{2009ApJ...697..573O}, (4) \citealt{2009ApJ...701.1076G}, (5) \citealt{2010ApJ...719L.109L},  (6) \citealt{2011ApJ...742...84O}, (7) \citealt{2014ApJ...790...29G}.}
  \end{tablenotes} 
\end {table*}

Rather than being acquired at its birth or posterior to it, the spin of the BH could be directly related to the AM of the progenitor star. \cite{2010Natur.468...77V} proposed a formation channel for the BH HMXB M33 X-7, where the initial binary has an orbit of a few days, and the BH progenitor star transfers part of its envelope to the secondary while still in the MS \citep[Case-A MT, see][]{1967ZA.....65..251K}. Assuming solid-body rotation during the MS phase and tidal locking while the binary is mass-transferring, the core of the BH progenitor contains large amounts of AM at the end of its MS phase. Having lost its envelope during the Case-A MT, the BH progenitor star never expands to become a giant star. Instead, after the end of the MS, it contracts to become a Wolf-Rayet star, and the binary remains in a close orbit of a few days during its whole lifetime. \cite{2012Sci...337..444S} found that $\sim 70\%$ of observed O-type stars are in close binary systems, and that half of these are close enough to undergo the Case-A MT, making this evolutionary path a common one. We also note that a series of systematic investigations \citep[][ and references therein]{2014ApJS..213...34K,2015ApJ...811...85K} of massive star binary characteristics in Cygnus OB2 associations have been carried out, which have slightly weaker constraints on the binarity due to limited observational samples.

For binaries close to Roche-lobe overflow at birth with sub-solar metallicities, enhanced rotational mixing has been predicted to result in the CHE of both stars \citep{2016MNRAS.458.2634M,2016A&A...585A.120S,2016A&A...588A..50M} or just the more massive component \citep{2009A&A...497..243D,2017A&A...604A..55M}. The latter case is realized in systems with initial mass ratios far from unity and results in the formation of BH HMXBs with high spins and a MS companion, providing an alternative channel to the Case-A MT. . 

In this Letter, we investigate the origin of the spin of the BH in HMXBs by studying in detail the evolution of close massive binaries, which leads to the Case-A MT and the CHE. The main methods used in the stellar and binary evolution models are discussed in \S2 and we present our results in \S3. We describe the resulting BH spins from the Case-A MT and the CHE in \S3.1, the relevant range in orbital periods, primary masses, and mass ratios leading to both formation channels in \S3.2, and also discuss how the Case-A MT leads to nitrogen enrichment of the BH companion in \S3.3. We then compare our results with current observations in \S4. Finally, the main conclusions of this Letter are summarized in \S5.

\section{Methods}
We use release 10398 of the MESA stellar evolution code \citep{2011ApJS..192....3P,2013ApJS..208....4P,2015ApJS..220...15P,2018ApJS..234...34P} to perform all of the binary evolution calculations presented in this Letter. We adopt a metallicity of $Z=Z_\odot/2$, where we take the solar metallicity to be $Z_\odot=0.017$ \citep{1996ASPC...99..117G}. The initial helium mass fraction is computed by assuming that it increases linearly from its primordial value of $Y = 0.2477$ \citep{2007ApJ...666..636P} at $Z = 0$ to $Y = 0.28$ at $Z = $Z$_\odot$. We model convection by using the standard mixing-length theory \citep{1958ZA.....46..108B} with a mixing-length parameter of $\alpha$ = 1.5 and adopt the Ledoux convection criterion. We model semiconvection according to \cite{1983A&A...126..207L} with an efficiency parameter of $\alpha_{sc}$ = 1.0. Step overshooting is considered with an extension given by $0.1H_P$, where $H_P$ is the pressure scale height at the convective core boundary. We model our binary systems until core carbon depletion in the center of the primary star.

Stellar winds are modeled following \cite{2011A&A...530A.115B}. For mass loss from hot hydrogen-rich stars ($X > 0.7$ at their surface) we use the prescription of \cite{2001A&A...369..574V}. For stars with a surface hydrogen of $X<0.4$, we use the mass-loss rate of \cite{1995A&A...299..151H} divided by a factor of 10 to account for clumping \citep{2010ApJ...725..940Y}. We further scale the mass-loss rate of \cite{1995A&A...299..151H} by a factor of (Z/Z$_{\odot})^{0.85}$, assuming the same metallicity dependence predicted by \cite{2001A&A...369..574V} for hydrogen-rich stars. We linearly interpolate these two mass-loss rates when the surface hydrogen $X$ is between $0.7$ and $0.4$.

We model rotational mixing and AM transport as diffusive processes \citep{2000ApJ...544.1016H}, including the effects of Eddington–Sweet circulations, the Goldreich–Schubert–Fricke instability, as well as secular and dynamical shear mixing. We also include diffusive element mixing from these processes with an efficiency parameter of $f_c = 1/30$ \citep{1992A&A...253..173C,2000ApJ...544.1016H}. For an efficient AM transport mechanism \cite[i.e., Tayler–Spruit dynamo;][]{1999A&A...349..189S,2002A&A...381..923S}, most of the internal AM is transported to the outer layers when the star leaves the MS.

Tides, in close binaries, play a critical role in the evolution of the orbit and the internal AM of the two stellar components. Here, we adopt the dynamical tide model derived by \citet{1975A&A....41..329Z}. The synchronization timescale, $T_{\rm sync}$, between the orbital period and the spin period of each star strongly depends on the tidal coefficient $\rm E_2$, which in turn depends on the structure profile of each stellar component. \cite{2018A&A...616A..28Q} recently computed $E_2$ for both H-rich and He-rich stars, in a wide range of initial masses, evolutionary stages, and at three different metallicities ($Z_{\odot}$, 0.1Z${_\odot}$, and 0.01Z${_\odot}$). For H-rich stars, the derived fitting formula relating the value of $E_2$ to the ratio of the convective core radius to the total radius of the star is given in Eq. 9 of that paper, and this is what we use throughout this Letter. In the standard implementation of tides in MESA, each layer of the star is synchronized independently on the timescale of $T_{\rm sync}$ \citep[i.e., equation (20) from][]{2015ApJS..220...15P}. Instead, in this Letter we implement a variation of that approach, where the tides operate only on the radiative layers. We have verified that this variation has a very small impact on our results. MT is treated as a conservative process, but as the accreting star is spun up due to accretion, enhanced stellar winds can lead to effectively fully non-conservative MT (see section 2.9 of \citealt{2015ApJS..220...15P} and references therein).
Relevant files to reproduce all of the calculations of this Letter can be found on the MESA website \footnote{http://cococubed.asu.edu/mesa$_{-}$market/inlists.html}.

\section{Results}
\subsection{Spin of BHs formed by the Case-A MT or the CHE}

 \begin{figure}[h]
     \centering
     \includegraphics[width=\columnwidth]{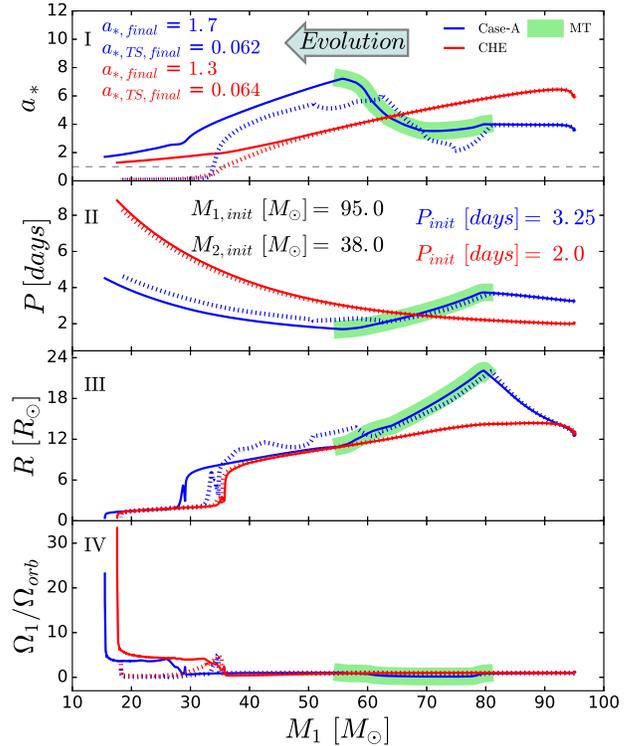}
     \caption{Spin parameter $a_*$ (I), orbital period (II), primary's radius (III), and rotational frequency ratio of primary to the orbit (IV) as a function of primary mass for two binary evolutionary sequences starting with same initial masses of two components but different initial orbital periods. The sequence with the longer initial period ($P_{init}=3.25$ days; blue line) evolves via the Case-A MT, while the one with the shorter initial period ($P_{init}=2.0$ days; red line) evolves via the CHE. Green shading represents the MT phase for the Case-A MT channel. The gray dashed line on the top panel indicates the theoretical maximum spin (i.e., $a_*$ = 1) of a BH and the arrow represents the direction of the evolution along the time. In both cases, assuming direct collapse, the BH progenitor star has enough AM to form a maximally spinning BH when it reaches core carbon depletion. For comparison, the dotted lines represent the same sequences but with an efficient AM transport mechanism.}
     \label{caseA-che}
\end{figure}

Here we investigate in detail the evolution of two close massive binaries that only have a different initial orbital period. In Fig.~\ref{caseA-che}, we show the evolution of various quantities including the spin parameter $a_*$ for the two representative binaries undergoing the Case-A MT and the CHE. The initial masses of the primary and the secondary, for both sequences, are 95.0 and 38.0 M$_{\rm \odot}$. For an initial orbital period of 3.25 days, the orbit initially expands to a period of about 4 days due to wind mass loss, at which point the primary star has expanded enough to fill its Roche lobe and initiate MT (shown in green shading). Since the binary is initially assumed to be synchronized, $a_*$ is already high ($\sim 3.8$) at the beginning of the simulation, and even increases slightly during the initial detached evolution, as the star expands during the MS increasing its moment of inertia. When the MT phase initiates, the primary star contracts due to mass loss in order to fit within its Roche lobe, and at the same time the orbit shrinks on a timescale of $\sim 1000$ years. These two processes have competing effects on the spin AM of the star. The decrease of the radius lowers the moment of inertia of the star, while the decrease of the orbital period increases the spin frequency of the star, which remains synchronized until the end of the MT phase. Overall, after an initial small decrease, $a*$ reaches its maximum value at the end of the MT phase. 

Shortly after the mass ratio of the binary reverses and the orbit starts expanding due to the MT, the binary detaches. The primary star continues to lose mass due to stellar winds, leading to orbital expansion and a gradual decrease of the spin parameter $a_*$. When the primary depletes hydrogen in its core, most of the hydrogen envelope has been lost and the entire star contracts until helium is ignited in its core. The timescale of contraction is much shorter than both the timescales of tidal synchronization and mass loss, so the star retains most of its AM, and loses corotation with the orbit; see panel (IV). The primary star, whose radius has now decreased by a factor of $\sim 5$, continues its evolution effectively as a single star, losing mass and AM only via stellar winds. Despite the intense mass loss, the primary star retains enough AM when it reaches core carbon depletion to form a maximally spinning BH.

Evolution is significantly different for a binary with the same component masses but a shorter orbital period (i.e., $P_{\rm init} = 2.0$ days). Enhanced rotational mixing leads to the CHE for the primary star, and its radius never expands to fill its Roche lobe. Instead, during its MS evolution the radius of the primary decreases due to stellar winds, and when core hydrogen is depleted, its radius quickly decreases by a factor of $\sim$ 4 as the star contracts to ignite helium. Since the binary never experiences Roche-lobe overflow, which would shrink the orbit, the final orbital period is larger than that of the Case-A MT sequence shown. The spin parameter $a_*$ of the primary is monotonically decreasing during the whole evolution and its final value is 1.3, retaining enough AM to form a fast-spinning BH.

We should note here that the efficiency of AM transport does not play a crucial role during MS evolution. Fig.~\ref{caseA-che} also shows the evolution of these two representative models including efficient AM transport from the Tayler–Spruit dynamo (see the dotted lines). We find that the evolution in the MS is similar, with both the Case-A MT and the CHE leading to the formation of a helium star with enough AM to produce a maximally spinning BH. The subsequent evolution, however, heavily depends on the AM transport efficiency, as tidal interaction becomes negligible and the star undergoes effectively single stellar evolution. Efficient AM redistribution coupled with strong wind mass loss rapidly depletes the AM of the whole star and our models that include the Tayler–Spruit dynamo result in BHs with spin parameters of $a_*<0.1$.

\subsection{Impact of the initial orbital period and primary mass on the various outcomes}

 \begin{figure}
     \centering
     \includegraphics[width=\columnwidth]{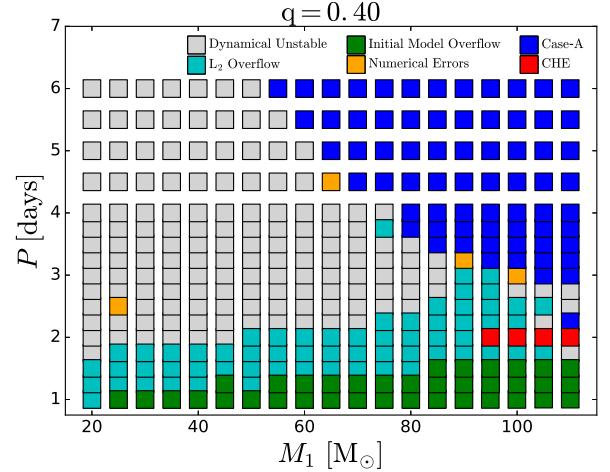}
     \caption{Outcomes of binary systems with a fixed mass ratio of $q=0.4$ and different initial orbital periods and primary star masses. The gray squares represent systems with MT rates higher than 10 $M_{\odot}$ $yr^{-1}$, which we consider as dynamically unstable, the cyan squares represent systems that overflow the second Lagrangian point $L_2$, and the green squares represent models that are overflowing at the zero-age MS (ZAMS). The blue squares represent systems that undergo the Case-A MT, red squares represent systems that undergo the CHE, and orange squares represent models with numerical errors and where the simulation was not completed.
     }
     \label{a_q4_noTS}
\end{figure}

 \begin{figure*}
     \centering
     \includegraphics[width=0.99\textwidth]{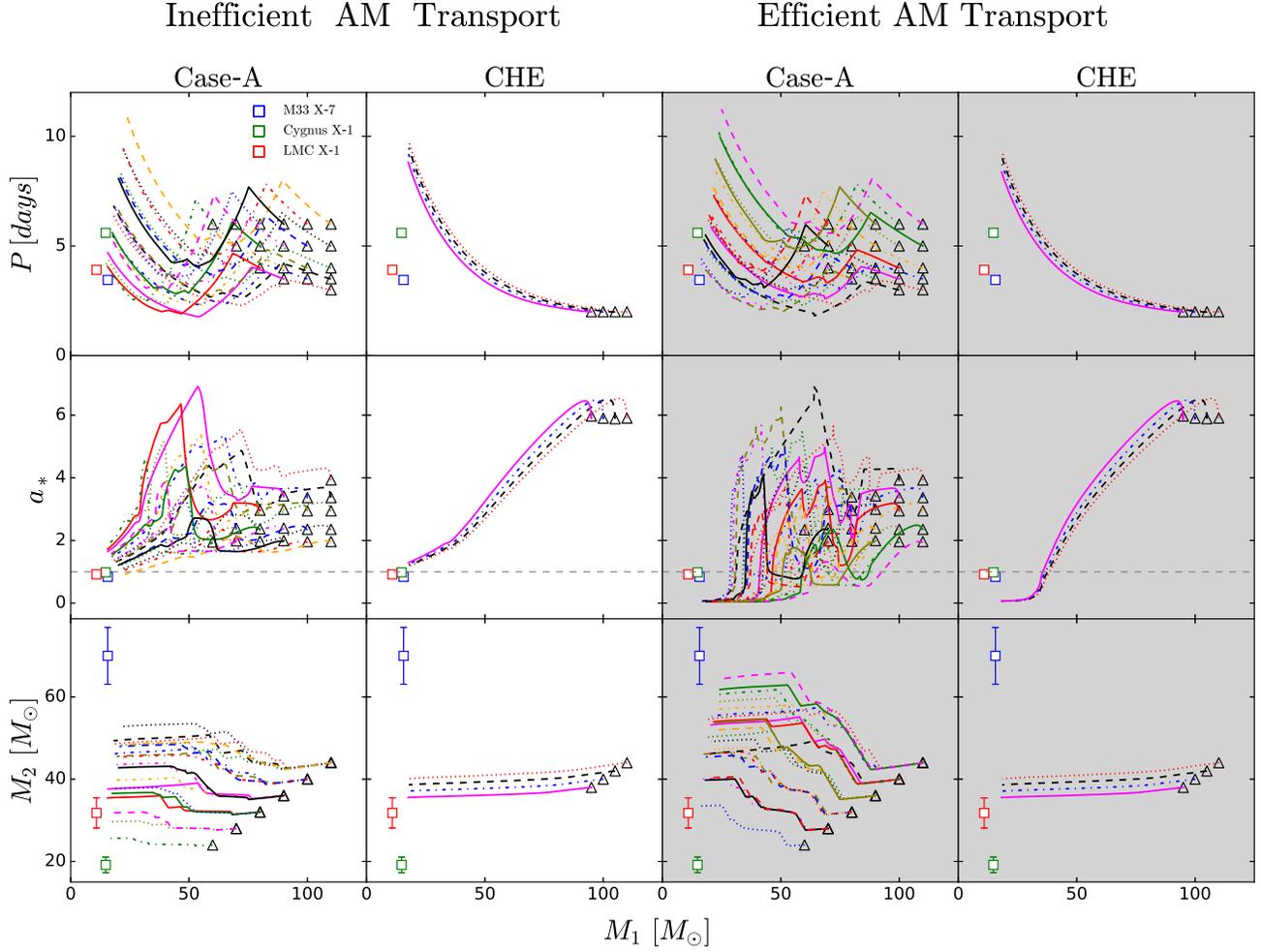}
     \caption{Evolution of the orbital period, spin parameter $a_*$ and secondary mass as function of primary mass for inefficient and efficient (namely with Tayler–Spruit dynamo marked by the gray background) AM transport mechanism. The black triangles refer to the initial properties of the binary systems, i.e., masses of two components and orbital period. In each column, one system is marked with the same color and line style. The squares show the properties of HMXBs with measured spins (blue for M33 X-7, green for Cygnus X-1, and red for LMC X-1). In the middle horizontal panels, the horizontal dashed line marks $a_*$ = 1.}
     \label{a_evol}
\end{figure*}

In order to explore the impact of the initial parameters on Case-A and CHE, we computed 4845 binary evolution sequences with varying primary star masses, mass ratios ($q=M_2/M_1$), and initial orbital periods. The primary masses range from 20 to 110 M$_{\odot}$ in intervals of 5 M$_{\odot}$, mass ratios from 0.25 to 0.95 in steps of 0.05, and initial orbital periods between 1 and 4 days in steps of 0.25 days and between 4 and 6 days with a lower resolution of 0.5 days. In Fig.~\ref{a_q4_noTS}, we show a slice of our grid with initial mass ratios of 0.4 (other mass ratios show qualitatively similar results). Our fiducial grid assumes inefficient AM transport. However, we repeated our calculations with the Tayler–Spruit dynamo operating in the interior of the star. Including the Tayler–Spruit dynamo does not change the outcomes shown in Fig.~\ref{a_q4_noTS} significantly, but alters the final BH spins dramatically.

For primary masses $M_1<60M_\odot$, most of the systems in Fig.~\ref{a_q4_noTS} undergo dynamically unstable MT and are expected to merge. The more massive primaries lose significant mass before the Roche-lobe overflow, reducing the mass ratio and leading to stable Case-A MT and the formation of a HMXB. Most binaries with initial orbital periods of $P<2$ days evolve into overcontact binaries extending beyond the second outer lagrangian point $L_2$ overflow \citep{2016A&A...588A..50M}, and are also expected to merge. When the initial orbital period becomes much shorter (i.e., $P_{\rm init} < 1.5$ days), the primary star overflows its Roche lobe at the ZAMS, representing a lower limit on the initial orbital period. Finally, CHE occurs only for a very small part of the parameter space, for orbital periods near overflow at ZAMS and high primary masses. This part of the parameter space has been shown to grow significantly for lower metallicities \citep{2017A&A...604A..55M}. Furthermore, here we point out that a convergence of $a_*$ to changes in spatial and temporal resolution was reached before running all of the simulations, which makes our result more reliable. The orange squares shown in Fig.~\ref{a_q4_noTS} correspond to the simulation that was not completed. Such numerical errors don not arise from some inadequacies in the code, but rather from the need to take very small time steps. Likely the proper handling of such situation would require a change in the numerical techniques.

Fig.~\ref{a_evol} shows the evolution of masses, orbital periods, and spin parameters $a_*$, for sequences from a slice of our grid with an initial mass ratio of 0.4. The systems that evolve via the Case-A MT channel (blue squares in Fig.~\ref{a_q4_noTS}) are shown on the first column of Fig.~\ref{a_evol}, where, for clarity, we only show half of the sequences. In the second column, all of the systems going through CHE (red squares in Fig.~\ref{a_q4_noTS}) are presented. In each column, black triangles refer to the initial conditions and the lines with same color and style correspond to the same binary system. The same grid is also calculated assuming efficient AM transport through the Tayler–Spruit dynamo, and the results are presented in the two columns with the gray background. We find that all of the primary stars in binary sequences with inefficient AM transport collapse to form BHs with high spins. In contrast, for all of the other systems with efficient AM transport mechanism, the BH spins are negligible. 

Finally, in each binary evolution, MT is initially treated as a conservative process. As the BH companion star is spun up, however, it reaches the critical rotation, which stops the accretion onto the secondary star, and the MT become non-conservative. In contrast, when non-conservative MT is initially assumed, a fast-spinning BH can still form, but more mass would be lost during the MT phase, which produces a wider binary system and hence a less massive BH companion. Overall, we expect that non-conservative MT throughout would just shift the properties of the progenitors that successfully match the observed systems.

\subsection{Enhancements of the nitrogen surface abundance via the Case-A MT channel}
\begin{figure}
     \centering
     \includegraphics[width=\columnwidth]{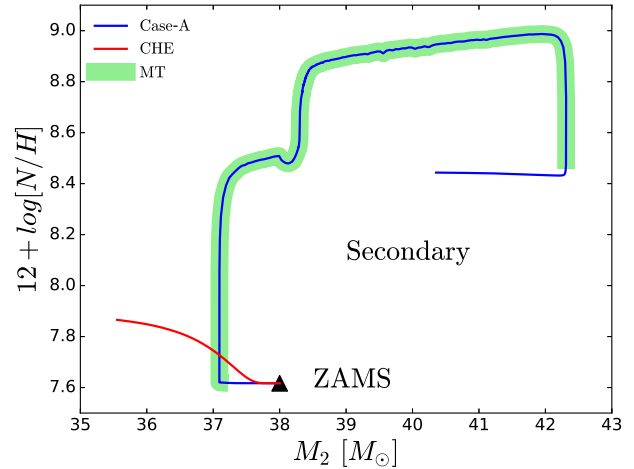}
     \caption{Nitrogen surface abundance of secondary as a function of its mass for the Case-A MT (blue solid line) and the CHE (red solid line), respectively. The green shading corresponds to the MT phase, which is same as that shown in Fig.~\ref{caseA-che}. The black triangle represents the ZAMS.}
     \label{abund}
\end{figure}

We also find that the Case-A MT leaves a distinct observational signature on the companion star, which could potentially allow us to distinguish them from HMXBs formed via the CHE or the classical common envelope evolution channel. In Fig.~\ref{abund}, we show the nitrogen surface abundance of the accreting star (which later becomes the donor during the HMXB phase) for the Case-A MT and the CHE sequences discussed in \S3.1. 
For the Case-A MT model, mass is transferred from deep layers of the primary that have been reprocessed from the CNO cycle and are thus nitrogen rich. This greatly enhances the nitrogen on the surface of the accretor \citep[see][]{2008IAUS..250..167L}. When MT stops, the nitrogen abundance drops due to dilution from thermohaline mixing, but its final value is still almost 1 dex above the pre-interaction value. In contrast, in the CHE channel much less important enhancements are reached, with $\sim0.3$ dex enhancement shown in Fig.~\ref{abund} arising from a combination of mass loss and mild rotational mixing.
In the classical common envelope channel, the two massive hydrogen-rich stars have an initially a wide orbit. When the primary star fills its Roche lobe in its giant phase, MT is dynamically unstable and the system undergoes a common envelope phase, during which the secondary is not expected to accrete any significant amount of mass. Thus, overall, no enhancements in the nitrogen surface abundance is expected. We then expect large ($\sim 1$ dex) enhancements of nitrogen abundance to be a characteristic property of the donor stars in BH HMXBs formed through the Case-A MT. 

\section{Comparison with observations}

Overall, Fig.~\ref{a_evol} shows that the CHE leads to final orbits that are too wide compared to the orbital periods of observed BH HMXBs. Furthermore, the parameter space at which the CHE occurs is very small compared to the parameter space corresponding to the Case-A MT, at least for the metallicities relevant to systems we consider here. For every BH HMXB originating from the CHE channel, one would expect to see many more coming from the Case-A MT. Both of these arguments point to the conclusion that the most likely formation channel for the three observed BH HMXBs with measured BH spins is the Case-A MT channel. 

In order to be more quantitative, we searched all of the sequences of our grid to find the ones that most closely resemble the observed properties of Cygnus X-1, LMC X-1, and M33 X-7 (see Table 1). The three best-fit sequences were selected by applying the minimum $\chi^2$ method to the observed properties (i.e., masses of the BH and its companion as well as the orbital period). For all three HMXBs,  0.5 days (the results are not sensitive to the choice.) is taken as the observational error of the orbital period to obtain the best match. Otherwise, the real observational error of orbital period is so small that its weight dominates the value of $\chi^2$.
Furthermore, we assumed that the BH was formed through a direct collapse, so the mass of the primary at central carbon exhaustion is equal to the mass of the resultant BH, and hence the orbital period remains unchanged after the BH formation. 

\begin{figure}
     \centering
     \includegraphics[width=\columnwidth]{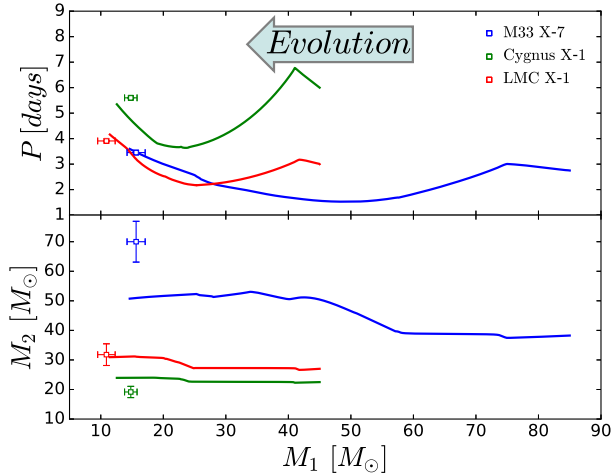}
     \caption{Orbital period (top panel) and secondary mass (bottom panel) as a function of primary mass. Blue, green and red solid lines correspond to the "best-fit" binary sequences that reproduce HMXBs that resemble M33 X-7, Cyguns X-1 and LMC X-1, respectively. The properties of the observed systems are marked with blue, red and green squares for M33 X-7, Cygnus X-1 and LMC X-1, respectively. The arrow on the top panel represents the direction of the evolution.}
     \label{match}
\end{figure}

In Fig.~\ref{match}, we show the three best-fit sequences and one can see that they match the BH masses and periods well. For LMC X-1, the selected sequences are consistent also with the companion mass.
For Cygnus X-1, we can see the mass of the companion star is about 1 $\sigma$ higher than the measured mass. A higher resolution of the grid might be required to better match it. Besides, surface abundance anomalies consistent with CNO processed material have already been observed in Cygnus X-1 \citep{2009ApJ...701.1895C}, providing additional support to the Case-A MT channel involving stable MT for this particular object. For the best-fit selected sequence of M33 X-7, the mass of the companion star is far below the measured value. This is because for the high initial primary mass and high initial mass ratio that are required in order to produce a system like M33 X-7, the companion star is being spun up due to accreted material, making the MT highly non-conservative. We should stress that although our prescription for the accretion efficiency is physically motivated, it remains approximate and highly uncertain. Had the MT been assumed to be conservative, as in \citet{2010Natur.468...77V}, the mass of the BH companion could reach much higher values.

\section{Conclusions}
In this Letter, we explore different AM transport mechanisms to investigate the AM of the BH progenitor via the Case-A and the CHE channels. We find that the efficiency of the AM transport does not play a crucial role during the MS phase. However, in order to form a fast-rotating BH in HMXB, weak coupling between the core and envelope inside the star after its MS phase is required both for systems evolving along the Case-A and the CHE channel.

The Case-A MT can explain the current properties of Cygnus X-1, LMC X-1, and M33 X-7 well. For the metallicity we have studied ($Z_\odot/2$), the CHE forms wider binary systems, which is not consistent with currently observed HMXBs with measured BH spins. The mismatch of the companion mass for M33 X-7 might be due to uncertainties in the prescription used here for the accretion efficiency, which requires further study. Furthermore, the Case-B channel, where MT is initiated after the primary depletes its central hydrogen, would result to an even wider HMXB orbit due to the longer initial period and earlier wind mass loss from the system, which makes such systems relatively dim \footnote{Based on the Bondi accretion mechanism \citep{1944MNRAS.104..273B}, the mean accretion rate was given in Equation \citep[6 of][]{2002MNRAS.329..897H}. For a very wide orbit (i.e., orbital velocity is far smaller when compared to the velocity of the donor's wind), an estimation can be obtained, namely $L_X \propto P_{\rm orb}^{-4/3}$, where $L_X$ is the luminosity in X-ray phase and $P_{\rm orb}$ is the orbital period. For Case-B or any other possible channels, the $P_{\rm orb}$ in the X-ray phase should be much larger, which results in a much more dimmer $L_X$.}. In contrast, the Case-A MT channel produces tight BH X-ray binaries with more massive donor stars, which makes such systems significantly bright and most likely to dominate the observed sample of BH wind-fed HMXBs. Quantitative predictions of the relative occurrence of each channel require population synthesis calculation, which will be the topic of a follow-up study.

Significant enhancements of the nitrogen surface abundance of donor stars in HMXBs can be produced in the Case-A MT channel. Thus it can be considered an important auxiliary tool to distinguish the Case-A MT channel from classical common envelope or the CHE channel. 

\begin{acknowledgements}
This work was sponsored by the Chinese Scholarship Council (CSC). This project has received funding from the European Union's Horizon 2020 research and innovation programe under the Marie Sklodowska-Curie RISE action, grant agreement No 691164 (ASTROSTAT). T.F. is grateful for support from the SNSF Professorship grant (project number PP00P2\_176868). P.M. and V.K. acknowledge support from NSF AST-1517753 and from a Senior Fellowship of the Canadian Institute for Advanced Research (CIFAR) program in Gravity and Extreme Universe. The computations were performed at the University of Geneva on the Baobab computer cluster. We would like to thank the anonymous referee for the helpful comments and suggestions.

\end{acknowledgements}


All figures were made with the free Python module Matplotlib \citep{2007CSE.....9...90H}.

\end{document}